\documentclass[useAMS]{mn2e}
\usepackage{psfig,epsfig}
\usepackage{graphicx}
\usepackage{amssymb}
\usepackage{epstopdf}

\begin{document}
\def\simlt{\mathrel{\rlap{\lower 3pt\hbox{$\sim$}}
        \raise 2.0pt\hbox{$<$}}}
\def\simgt{\mathrel{\rlap{\lower 3pt\hbox{$\sim$}}
        \raise 2.0pt\hbox{$>$}}}

\title[Environment of radio AGN]{The environmental properties of radio-emitting AGN}

\author[Manuela Magliocchetti et al.]
{\parbox[t]\textwidth{M. Magliocchetti$^{1}$, P.Popesso$^{2}$, M. Brusa$^{3,4}$, M. Salvato$^{5}$\\
} \\
{\tt $^1$ INAF-IAPS, Via Fosso del Cavaliere 100, 00133 Roma,
  Italy}\\
 {\tt $^2$ Excellence Cluster, Boltzmannstr. 2, D85748, Garching, Germany}\\
  {\tt $^3$ Dipartimento di Fisica e Astronomia, Universita' di Bologna, Via Gobetti 93/2, 40129, Bologna, Italy}\\
  {\tt $^4$ INAF-Osservatorio Astronomico di Bologna, Via Gobetti 93/3, 40129, Bologna, Italy}\\
  {\tt $^5$ Max Planck Institut f\"ur extraterrestrische Physik (MPE),  Postfach 1312,  D85741, Garching, Germany}\\
  }
 \maketitle
 \begin{abstract}
 We study the environmental 
properties of $z\simlt 1.2$ radio-selected AGN belonging to the $\sim 2$ deg$^2$ of the COSMOS field, 
finding that about 20\% of them 
appear within overdense structures. AGN with $P_{1.4 \rm GHz}>10^{23.5} \rm W  Hz^{-1}  sr^{-1}$ are twice more likely to be found in clusters 
with respect to fainter sources ($\sim 38$\% vs $\sim 15$\%), just as radio-selected AGN with stellar 
masses $M_*>10^{11} M_\odot$ are  twice more likely to be found in overdense 
environments with respect to objects of lower mass ($\sim 24$\% vs $\sim 11$\%). 
Comparisons with galaxy samples further suggest that radio-selected AGN of large stellar mass tend to avoid underdense environments more than normal galaxies with the same stellar content.
Stellar masses also seem to determine the location of radio-active AGN within 
clusters: $\sim 100$\% of the sources found as satellite galaxies have  $M_*<10^{11.3} M_\odot$, while $\sim 100$\% of the AGN coinciding with a cluster central galaxy have $M_*>10^{11} M_\odot$. No different location within the cluster is instead 
observed for AGN of various radio luminosities.  Radio AGN which also emit in 
the MIR show a marked preference to be found as isolated galaxies ($\sim 70$\%) at variance with those also active in the X-ray which all seem to reside within overdensities. 
What emerges from our work is a scenario whereby physical processes on sub-pc and kpc scales (e.g. emission respectively related to the AGN and to star formation) 
are strongly interconnected with the large-scale environment  of the AGN itself.
 \end{abstract}

\begin{keywords}
cosmology: dark matter - cosmology: large-scale structure of Universe - cosmology: observations - 
galaxies: starburst - galaxies: active - 
radio continuum: galaxies 
\end{keywords}
 
\section{Introduction}
Clustering studies have ubiquitously demonstrated that  AGN which are active at radio wavelengths (hereafter radio-active AGN or simply radio AGN) tend to occupy dark matter halos which are at least a factor 10 more massive than those inhabited by radio-quiet AGN. Furthermore, such values for the halo masses ($M_{\rm DM}\simgt 10^{14} M_\odot$) are comparable with those of structures hosting rich groups or clusters of galaxies 
(e.g. Magliocchetti et al. 2004; Porciani, Magliocchetti \& Norberg 2004; Best 2004; Wake et al. 2008; Shen et al. 2009; Fine et al. 2011; Lindsay et al. 2014a; Lindsay, Jarvis \& McAlpine 2014b;  Malavasi et al. 2015; Magliocchetti et al. 2017; Retana-Montenegro \& R\"ottgering 2017).

Indeed, radio AGN at all redshifts have been widely used as beacons to search for overdense structures (for some of the most recent results see e.g. Wylezalek et al. 2013; Castignani et al. 2014; Hatch et al. 2014; Shen et al. 2017 and references therein).

However, to our knowledge, none of the previous works tackle the more general issue of occurrence of radio-emitting AGN within cosmological environments of different kinds, especially once we move away from the local Universe. 
\begin{figure*}
\includegraphics[scale=0.42]{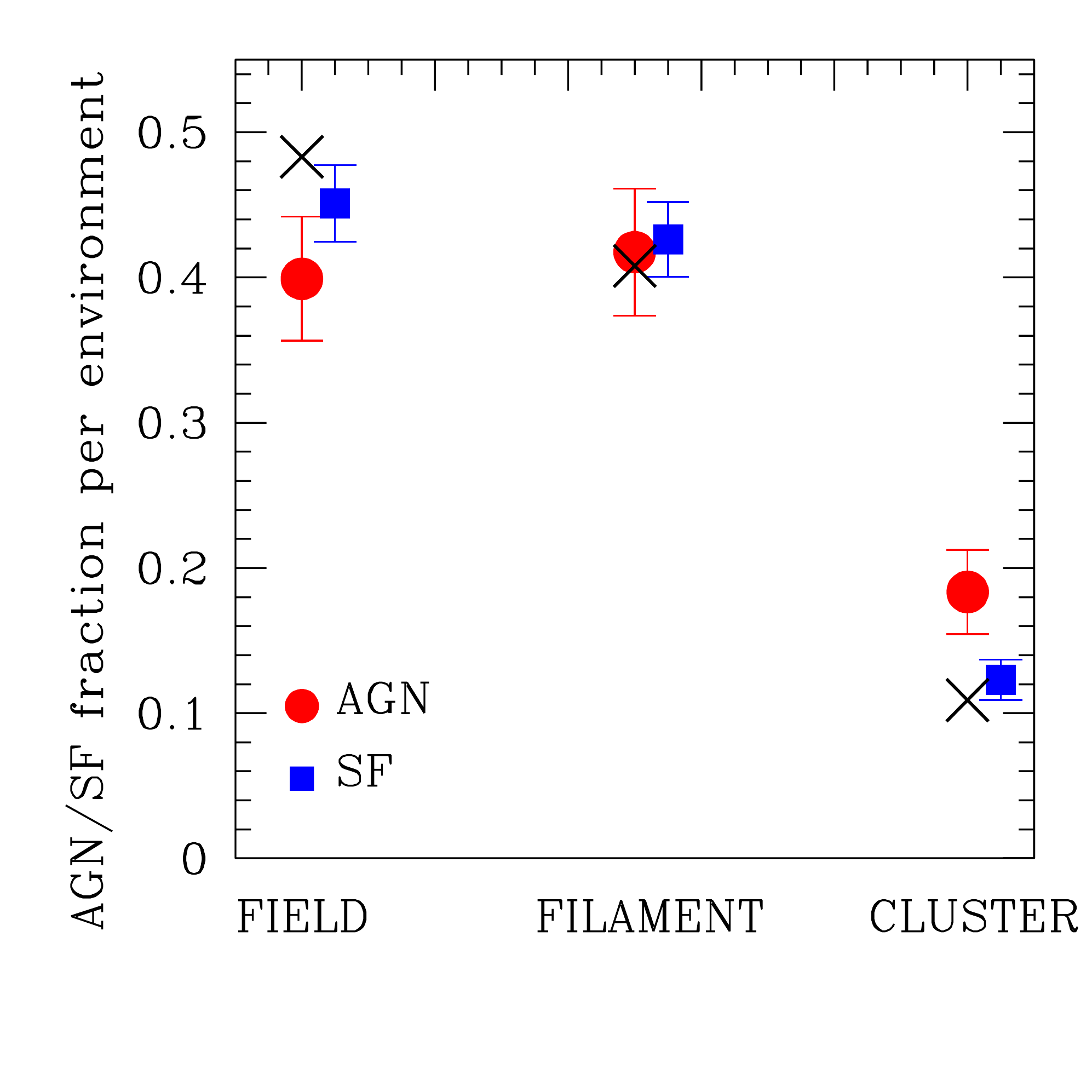}
\includegraphics[scale=0.42]{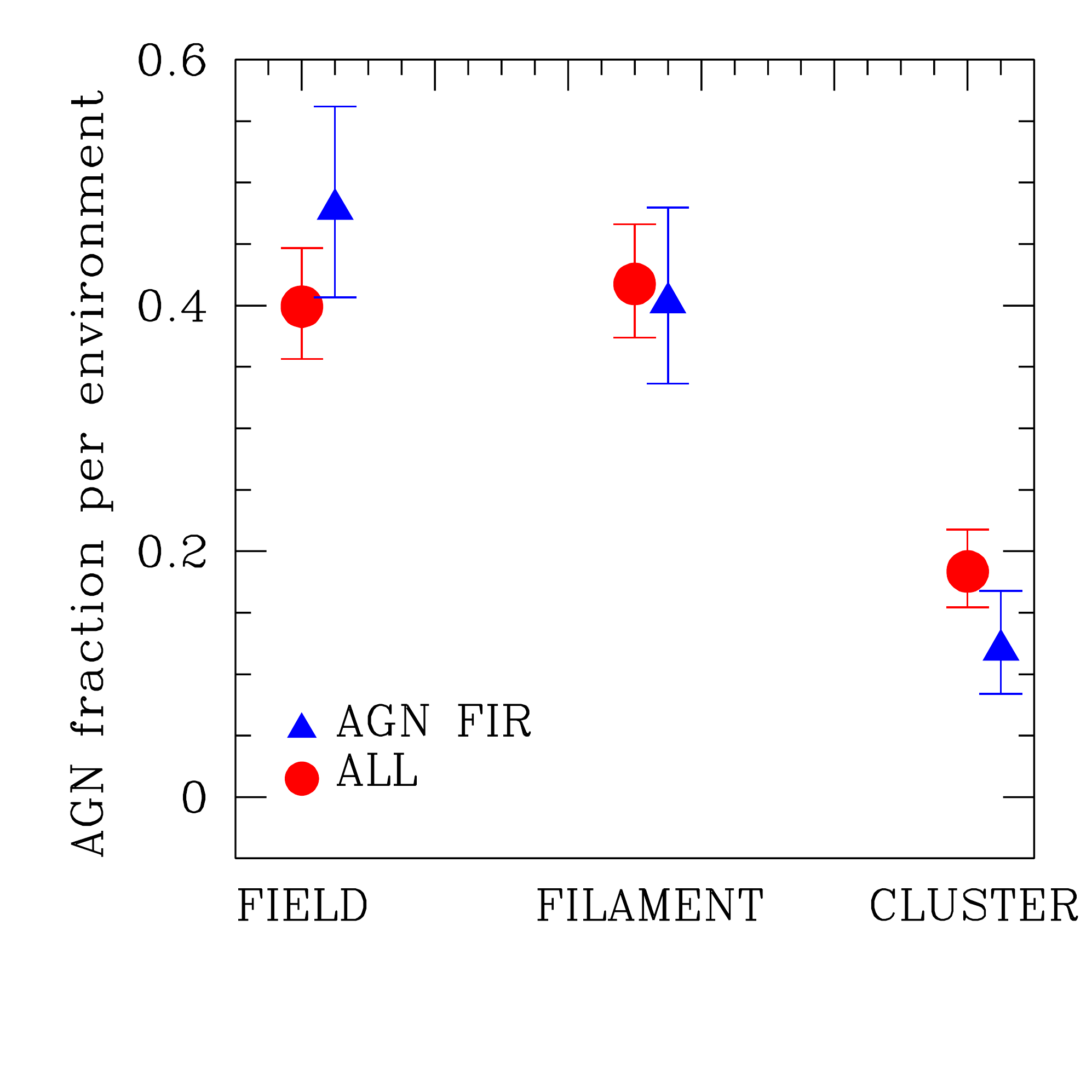}
\caption{Left-hand panel: fraction of radio-selected AGN (red dots) and star-forming galaxies (blue squares) within different environments. Crosses represent the results obtained by Darvish et al. (2017) for the whole population of COSMOS galaxies.  Right-hand panel: comparison between the environmental properties of radio-selected AGN (red dots) and those of the sub-population of radio AGN associated to ongoing star formation (blue triangles). For the sake of clarity, data points referring to star-forming galaxies in the left-hand panel and to FIR-selected AGN in the right-hand panel have been slightly shifted towards the right. In both plots error bars show 1$\sigma$ Poisson uncertainties; in the case of the Darvish et al. (2017) results, these are smaller than the symbols themselves.
\label{fig1}}
\end{figure*}	
Our paper aims to fill this gap by presenting a direct approach which pinpoints radio AGN within various cosmological structures such as clusters, filaments and the field. These structures were singled out up to $z=1.2$ by Darvish et al. (2017) in that portion of the sky which corresponds to the COSMOS survey (Scoville et al. 2007). Furthermore, the wealth of data available for galaxies in the COSMOS field also allows us to investigate the environmental properties of radio AGN as a function of different physical parameters such as  radio luminosity, AGN output at other wavelengths, star-formation activity and stellar mass of their hosts. This in turn allows us to draw important conclusions on the possible connections between processes happening 
on extremely different (from sub-pc up to a few Mpc) physical scales and to shed light on the feedback mechanisms which are triggered by radio activity of AGN origin. 



\section{The catalogue}
The main properties of the population of radio-selected AGN  considered in the present work are extensively described in Magliocchetti et al. (2014;  2018). 
Briefly, they were drawn from the original catalogue of 2382 radio objects brighter than a 1.4 GHz integrated flux density of 60 $\mu$Jy stemming from the VLA-COSMOS Large Project  (Schinnerer et al. 2007; Bondi et al. 2008)\footnote{http://irsa.ipac.caltech.edu/data/COSMOS/tables/vla/vla-cosmos\_lp\_sources\_v2\_20080615.tbl}. 
In order to find redshifts for as many sources as possible, Magliocchetti et al. (2018) cross-correlated the above sample with the Laigle et al. (2016)  catalogue of photometric redshifts and stellar masses. 
The above procedure provided redshift estimates for 2123 radio sources, corresponding to $\sim 90$\% of the parent sample.
1173 of them were also found to have a counterpart in the {\it Herschel} catalogues released by the PACS Evolutionary Probe (PEP,  D. Lutz et al. 2011) team either at 100$\mu$m or at 160$\mu$m. 

Magliocchetti et al. (2018) used the crossover points of the luminosity functions for AGN and star-forming galaxies in Figure 7 of McAlpine, Jarvis \& Bonfield (2013) to classify an AGN if its radio luminosity (calculated assuming a radio spectral index $\alpha=0.7$  for $F_\nu\propto \nu^{-\alpha}$ and a $\Lambda$CDM cosmology with $H_0=70 \: \rm km\:s^{-1}\: Mpc^{-1}$ and $\Omega_0=0.3$) was higher than some threshold: $\rm Log_{10}P_{\rm c}(z)=\rm Log_{10}P_{0,\rm c}+z$ for $z\le 1.8$ and $\rm Log_{10}P_{\rm c}(z)=23.5$ $\rm W Hz^{-1} sr^{-1}$ at higher redshifts. Otherwise the galaxy is classified as star forming. $P_{0,\rm c}=10^{21.7}$ $\rm W Hz^{-1} sr^{-1}$ holds in the local universe and  roughly coincides with the break in the radio luminosity function of star-forming galaxies (e.g. Magliocchetti et al. 2002). The above choice arises naturally, since beyond $\rm P_{c}$  the luminosity function of star-forming galaxies steeply declines, and their contribution to the total radio population is drastically reduced to a negligible percentage.
This procedure identifies 704 AGN (corresponding to 33\% of the whole radio population) and 1419 star-forming galaxies.  272 sources classified as AGN and 901 sources classified as star-forming galaxies were also found in the {\it Herschel} catalogues.

We then considered the catalogue of environments produced by Darvish et al. (2017) by using all those galaxies within the COSMOS field which have stellar masses $M_*\ge 9.6$ M$_\odot$  (Chabrier IMF), photometric redshifts (as taken from the Laigle et al. 2016 work) 
$0.1\le z\le 1.2$ and are located within the UltraVISTA-DR2 region (Ilbert et al. 2013) defined by $149.33^\circ\le\alpha(2000)\le 150.77^\circ$ and $1.615^\circ\le\delta(2000)\le 2.80^\circ$. These authors reconstruct the density field as traced by $0.1\le z\le 1.2$ galaxies and from it they extract the components of the cosmic web such as filaments, clusters and the field so that each considered galaxy is associated to its environment. 
Our radio-selected objects were then paired to those of Darvish et al. (2017) whenever their angular distance was within 1 arcsec. 
By doing this, we end up with 218 radio-selected AGN and 645 radio-selected star-forming galaxies with precise photometric redshift determinations ($\sigma_{{\Delta z}/(1+z)}\simlt 0.015$; cfr also Figure 2 from Darvish et al. 2017) and detailed information on their environmental properties. This will be our working sample. Note that whenever throughout the paper we will use the expression "overdense environments" we will only refer to those structures identified by the Darvish et al. 2017 work as "clusters".

Following Magliocchetti et al. (2018), we further flag these sources as X-ray AGN if they were listed as such in Brusa et al. (2010) and Marchesi et al. (2016) and MIR AGN if they are candidate AGN either detected by MIPS at 24$\mu$m and with a Spectral Energy Distribution (SED) in the IRAC bands presenting a power-law trend (Chang et al. 2017), or defined by a high signal-to-noise power-law behavior  in the IRAC bands (Donley et al. 2012). The breakdown is presented in Table 1, while the properties of all  individual sources are presented in Table 2.


\section{Results}
\subsection{Connections with the host galaxy properties}
\begin{figure}
\includegraphics[scale=0.42]{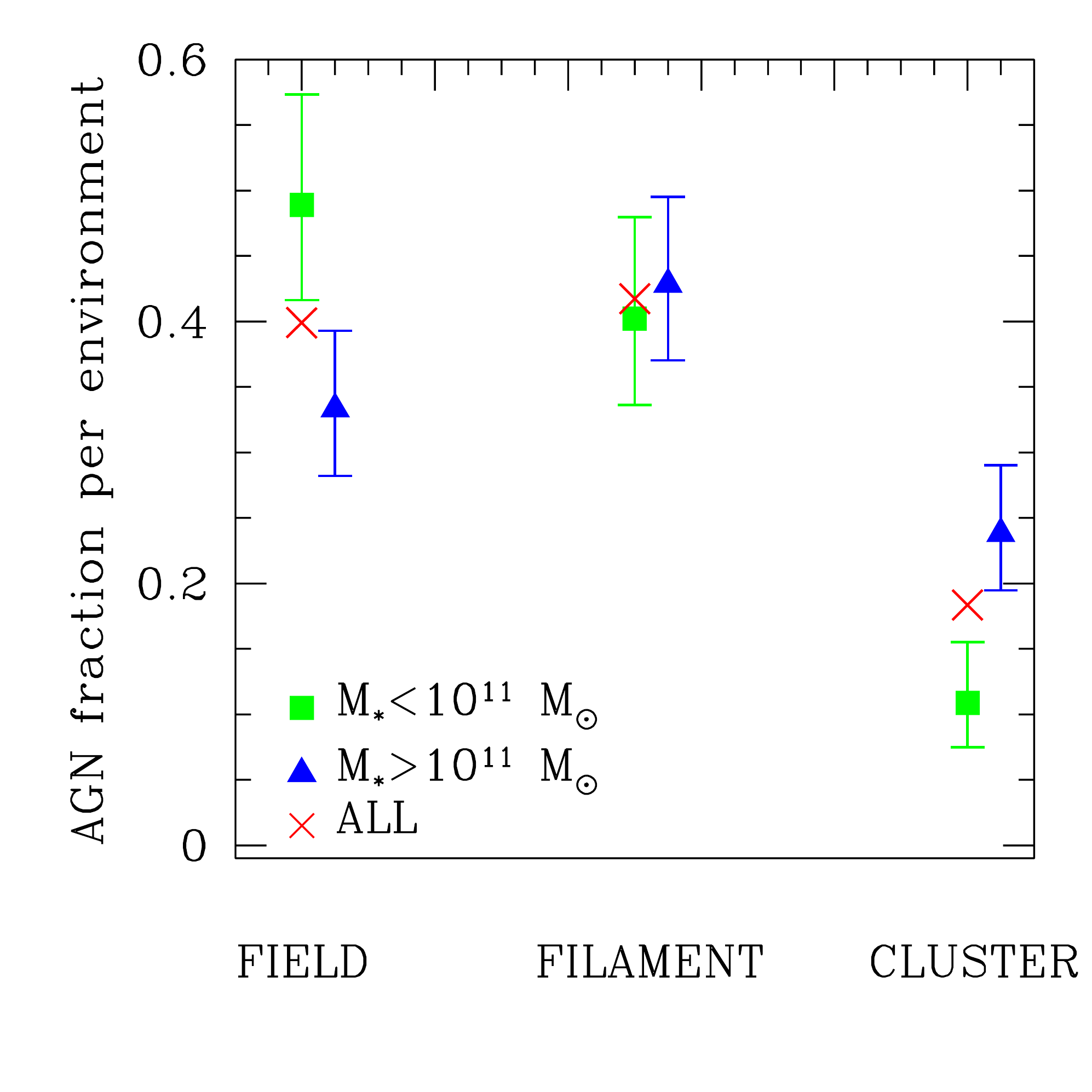}
\caption{Fraction of radio-selected AGN within different environments for different values of the stellar mass of the hosts. For the sake of clarity, data points referring to galaxies with stellar masses $M_*\le10^{11} M_\odot$ have been slightly shifted to the right. Error bars show 1$\sigma$ Poisson uncertainties. \label{fig2}}
\end{figure}
\begin{figure}
\includegraphics[scale=0.42]{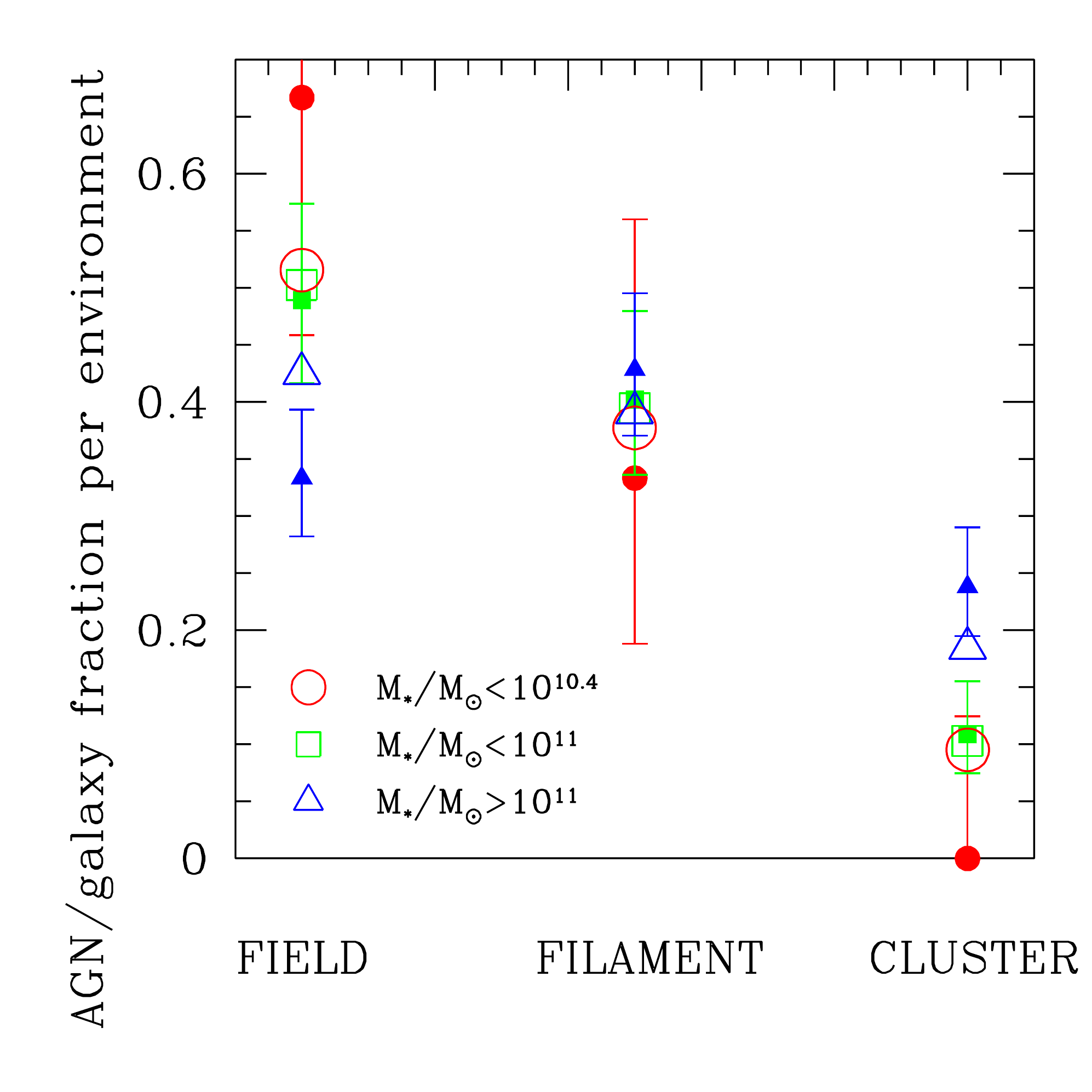}
\caption{Fraction of radio-selected AGN from this work (small filled symbols) and galaxies from Darvish et al. (2017) (large open symbols without error bars) within different environments for various values of the stellar mass of the hosts. 
 Circles indicate the results obtained for sources with stellar masses $M_*/M_\odot\le10^{10.4}$, squares for sources with $M_*/M_\odot <10^{11}$ and triangles for  $M_*/M_\odot\ge10^{11}$.
Error bars show 1$\sigma$ Poisson uncertainties. Those corresponding to the galaxies from Darvish et al. 2017 are smaller than the symbol sizes.} \label{fig3}
\end{figure}

As a first test, we compare the environmental properties of the two populations of radio-selected AGN and star-forming galaxies. 
This is done in the left-hand panel of Figure \ref{fig1} which shows the fractions which are respectively found in the field, in filaments and in clusters. 
As it is possible to notice, while both populations substantially occupy  filamentary structures in the same way, the abundance of radio-selected AGN ($18\pm 3$\%)  in overdense environments  is higher 
than those of both radio-selected star-forming galaxies ($12\pm 1$\%) and the general population of COSMOS galaxies (10.9\%, Darvish et al. 2017). On the other hand, 
although the present data does not allow to draw any significant conclusion, there seems to be a hint for a higher occurrence of star-forming galaxies in the field with respect to AGN.
The fact that, at least in the nearby universe, star-forming galaxies are more isolated than AGN which instead tend to inhabit richer environments is not a new result (cfr \S 1). 
It is however interesting to confirm this behavior also at radio wavelengths, although with a low statistical significance. 

As already discussed in Magliocchetti et al. (2018), the population of star-forming galaxies selected in \S 2 is  a mixed bag of all those sources which emit in the radio band thanks to processes connected to stellar formation. These include genuine star-forming galaxies as well 
as radio-quiet AGN whose host galaxies are actively forming stars (e.g. Padovani et al. 2015). It follows that it is quite difficult to draw conclusions on such a heterogeneous mix of sources, except for the observational fact that they 
trace the spatial distribution of more "normal" galaxies  (cfr left-hand panel of Figure 1, where the black crosses illustrate the results of Darvish et al. 2017 for the more general population of COSMOS galaxies). This is  why in the following we will only concentrate on {\it bona-fide} radio AGN (selected as in \S 2) which owe their emission to accretion onto a central black hole. 
  
We then investigate whether there is a connection between the presence of ongoing star formation within the AGN hosts and the environment within which these sources reside. 
As shown by the right-hand panel of Figure 1 which illustrates the distribution of radio-selected AGN also detected in the PACS-{\it Herschel} catalogues at Far-Infrared wavelengths compared to that of the whole radio AGN population, no appreciable difference is found amongst the two populations. The data at hand therefore suggests that 
there is no connection between the star-forming activity of a galaxy hosting a radio-active AGN and its large-scale environment.

\begin{figure}
\includegraphics[scale=0.42]{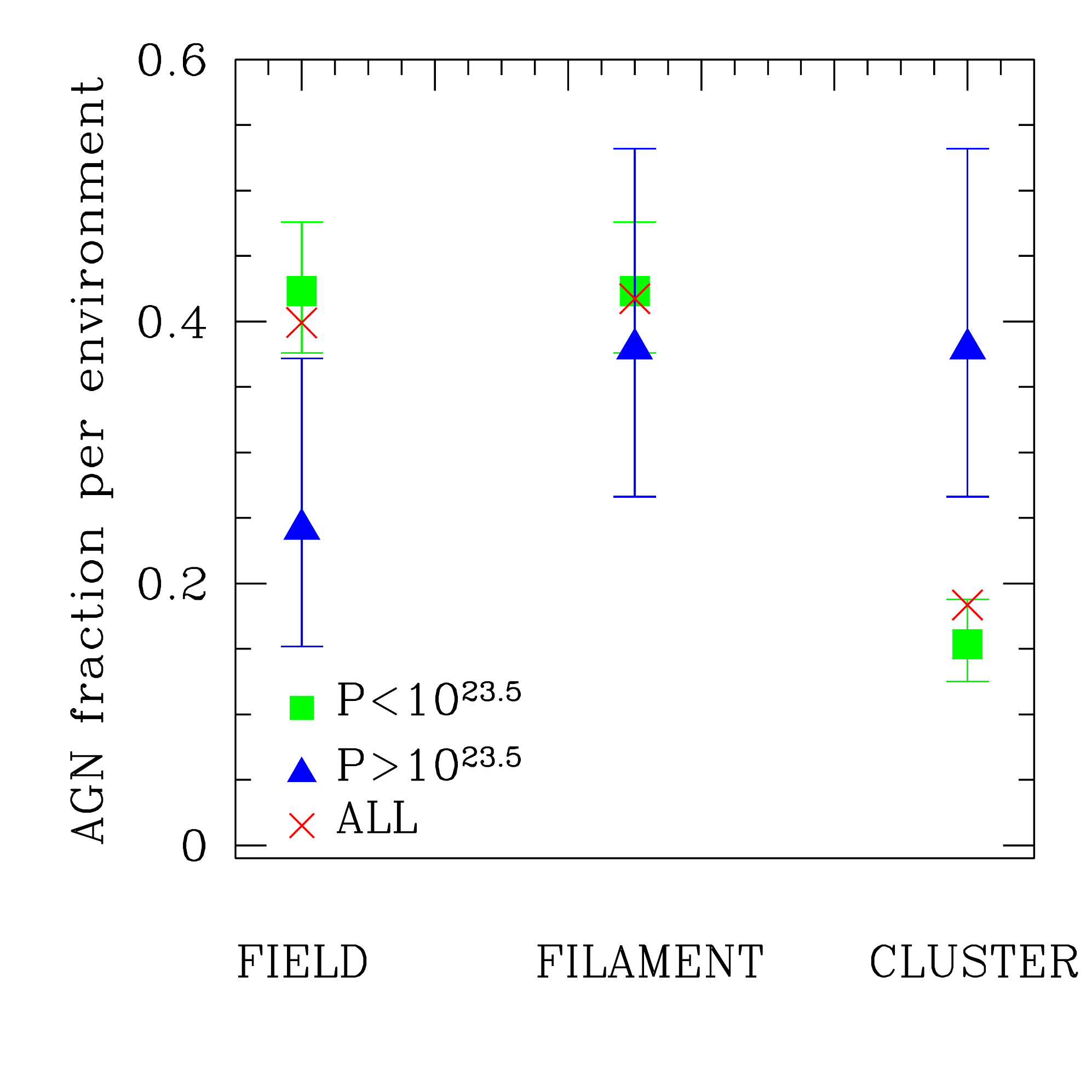}
\caption{Distribution of radio-selected AGN within different environments for different values of the  AGN radio luminosity (expressed in $\rm W Hz^{-1}sr^{-1}$). Error bars show 1$\sigma$ Poisson uncertainties. \label{fig4}}
\end{figure}	
 \begin{figure}
\includegraphics[scale=0.42]{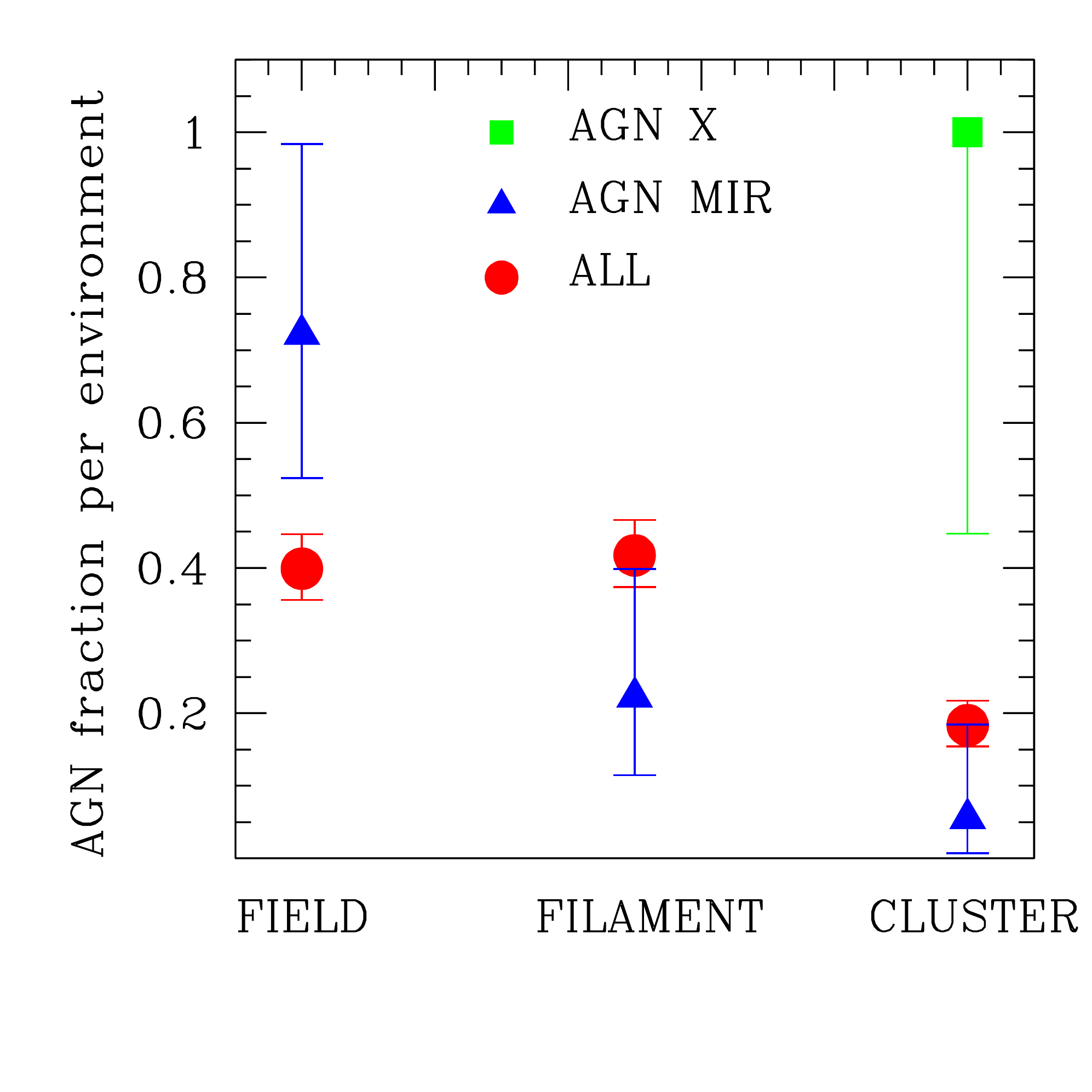}
\caption{Fraction of radio-selected AGN which also emit in other wavebands as a function of environment. Error bars show 1$\sigma$ Poisson uncertainties. \label{fig5}}
\end{figure}

More interestingly, we can investigate how radio-selected AGN with different physical properties occupy different environments. Indeed, we find that AGN hosted by galaxies with a large stellar mass content are about twice as likely to be found within rich environments with respect to those associated 
with hosts of smaller stellar masses.  In more quantitative terms, we have that $24^{+5}_{-4}$\% of radio-active AGN with $M_*\ge10^{11}$ M$_\odot$ are found within clusters, while this is true for only  $11^{+4}_{-3}$\% of the sources with $M_*<10^{11}$ M$_\odot$ (cfr Figure \ref{fig2}). We stress that such a decrement is balanced off only in the field as there are more low-mass radio-selected AGN in underdense environments ($49^{+8}_{-7}$\%) with respect to more massive ones ($33^{+6}_{-5}$\%), while we observe no difference in the environmental properties of the two sub-populations within filaments. We note that our findings confirm those obtained in the local universe (e.g. Magliocchetti \& Br\"uggen 2007; Best et al. 2007), and extend them to $z\sim 1.2$.

The results presented in Figure \ref{fig2} are not only due to the effect of mass segregation applied to radio-emitting AGN. We show this point in Figure \ref{fig3} which illustrates the fraction of radio-selected AGN from this work (small filled symbols) compared to that of galaxies from Darvish et al. (2017, large empty symbols) found within different environments for different values of the stellar mass of the hosts. 
The three considered ranges are $ M_*/M_\odot\le10^{10.4}$ (circles), $M_*/M_\odot<10^{11}$ (squares) and  $M_*/M_\odot\ge10^{11}$ (triangles). 
As Figure 3 suggests, galaxies and radio AGN of similar mass behave rather differently in terms of their environmental properties. 
In fact, we find a  $\sim 2\sigma$ indication for a decrement of massive, $M_*>10^{11}$ M$_\odot$,  radio-selected AGN in underdense cosmological structures such as the field in comparison with massive normal galaxies ($33^{+6}_{-5}$\% vs $43\pm 1$\%). 
One the other hand, there also seems  to be a trend for high-mass radio-selected AGN to be more abundant within cluster environments with respect to massive "normal galaxies" ($24^{+5}_{-4}$\% vs  $18.4\pm 0.8$\%), although this result is at a lower significance level. Once again, no appreciable difference is observed within filaments.

Interestingly enough, radio-selected AGN and normal galaxies of intermediate, $M_*/M_\odot<10^{11}$, stellar masses seem to show the same behavior at all cosmological scales. However, if we restrict our attention to sources with $M_*/M_\odot\le10^{10.4}$, it appears a tendency for radio-emitting AGN associated to small stellar mass systems to avoid overdense environments more than the more general population of galaxies with the same stellar mass content. 
In quantitative terms,  we have that $9.6\pm 0.2$\% of the galaxies with stellar masses $M_*/M_\odot\le10^{10.4}$ are found within cluster environments, while there is no radio-selected AGN (0 out of 15) in the same environment and mass range. Although masked by the large uncertainties, the above trend seems to be balanced in the field, whereby we find more low-mass radio-active AGN (10 out of 15, corresponding to $\sim 67$\% of the sub-sample) than low-mass galaxies ($52.2\pm0.4$\%).

The above discussion suggests that the environmental properties of radio-selected AGN may not be simply  dictated by mass segregation effects arising from the fact that radio-emitting sources tend to inhabit more massive galaxies which in turn tend to preferentially occupy denser structures. In fact, on top of this effect which certainly dominates the present case, we observe a $(1.5-2)\sigma$ difference in the environmental behaviors of radio AGN and normal galaxies with a similar stellar content. 
The above results are however hampered by the low counting statistics in our radio AGN sub-samples (few tens of AGN) and need to be  confirmed by wider/new generation radio surveys. If confirmed, the observed differences can almost certainly be ascribed to the presence of nuclear radio activity. Within this framework, our results suggest that such nuclear radio activity tends to be favoured either in large,  $M_*\ge10^{11}$ M$_\odot$, stellar mass systems residing within overdense structures or in smaller ($M_*/M_\odot\le10^{10.4}$) isolated galaxies. We note that Tasse et al. (2008) reach a very similar conclusion by analyzing a sample of 110 radio-selected AGN drawn from the XMM-LSS survey region. These authors indeed find that high-mass ($M_*\simgt 10^{10.5-10.8}$ M$_\odot$) radio-active galaxies prefer large-scale overdense regions, while the opposite is true for lower-mass AGN. They further find that these two classes of sources have different cosmological evolutions and speculate on the possibility of two well distinct triggering mechanisms 
(cooling of hot gas within the cluster environment for $M_*\simgt 10^{10.5}$ M$_\odot$, galaxy mergers and interactions for $M_*\simlt 10^{10.5}$ M$_\odot$)
for nuclear radio activity taking place in the low-mass and high-mass regime.  

\begin{figure*}
\includegraphics[scale=0.28]{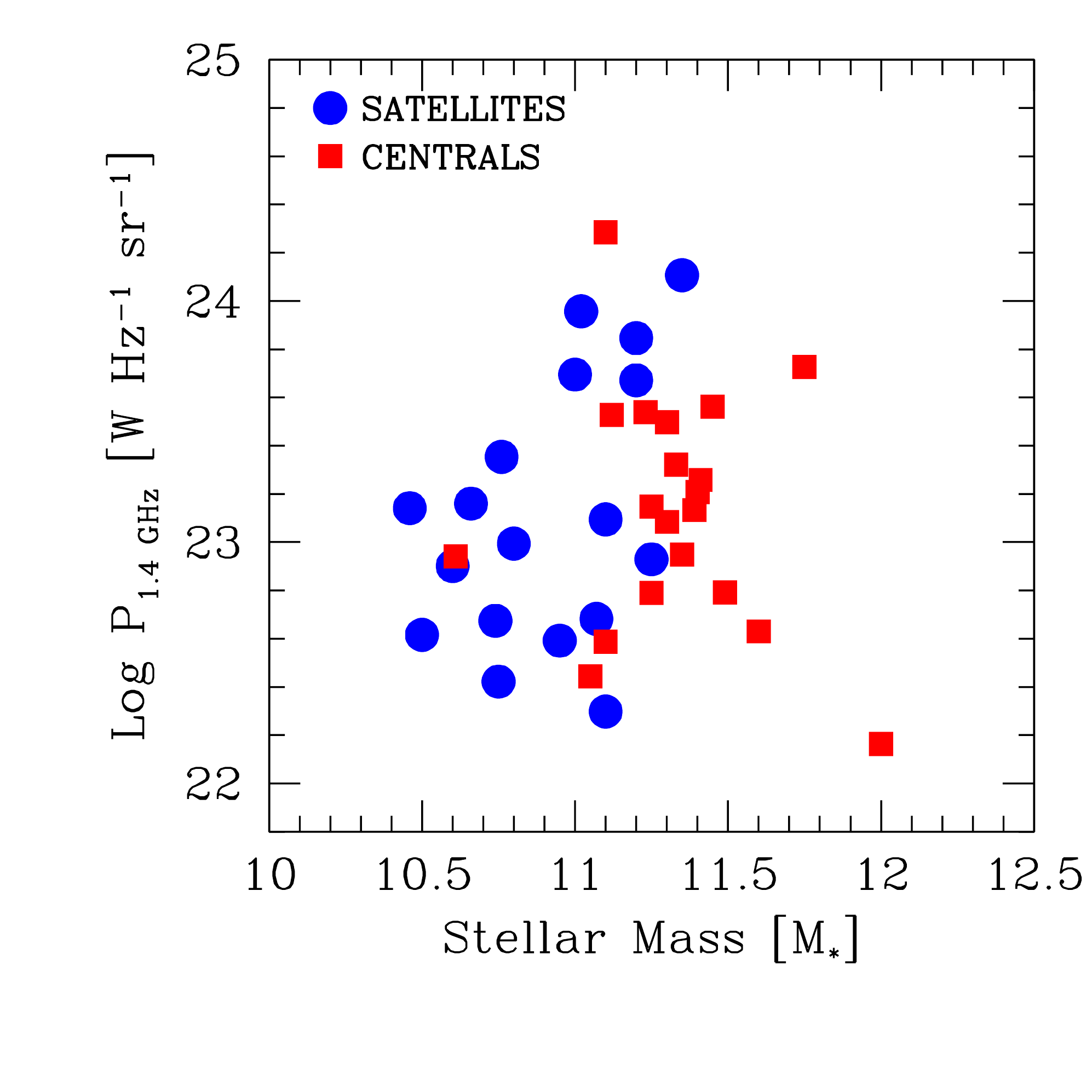}
\includegraphics[scale=0.28]{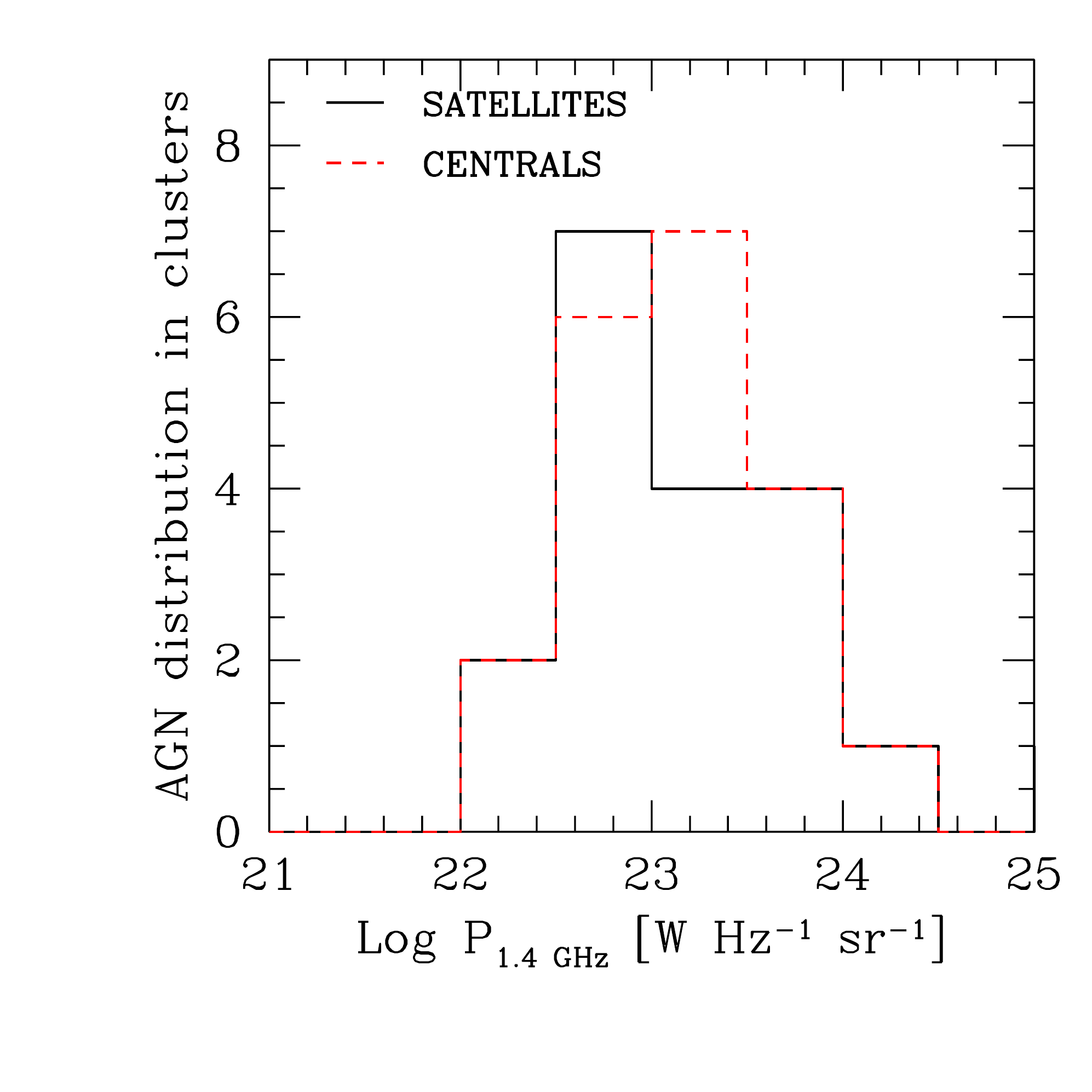}
\includegraphics[scale=0.28]{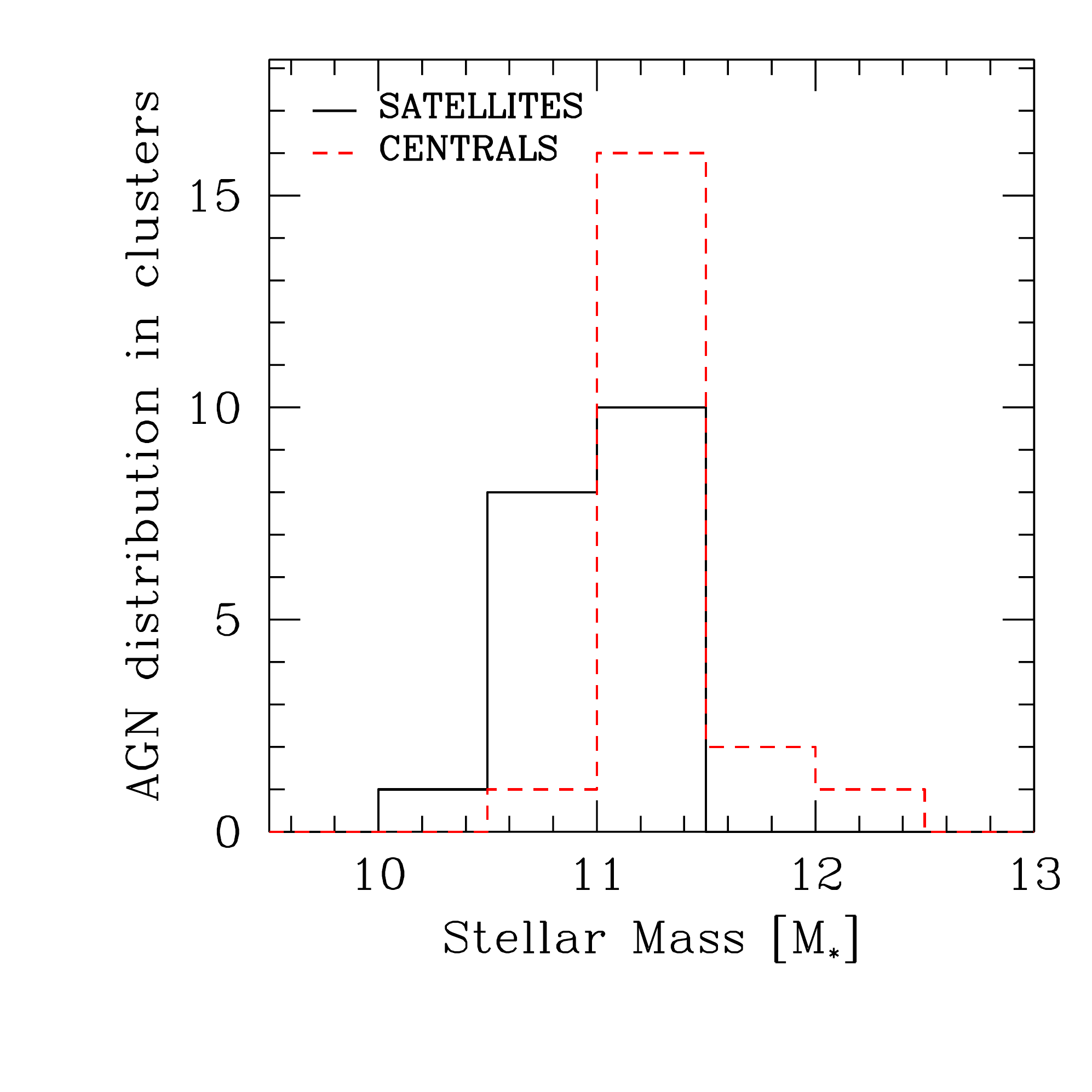}
\caption{Distributions of stellar masses and radio luminosities for radio-selected AGN identified either as satellite galaxies or as the central one of the cluster environments they belong to. \label{fig6}}
\end{figure*}	

\subsection{Connections with nuclear activity}
The issue of a possible connection between nuclear radio luminosity in radio-selected AGN and the environment they reside in has been recently matter of debate. 
Just to mention a few works, Donoso et al. (2010) analysed $\sim 14000$ radio-active AGN in the redshift range $0.4<z<0.8$ and found a strong positive trend between radio luminosity and environment for  P$_{1.4\rm GHz}\simgt 10^{23.5}$ ${\rm W Hz^{-1} sr^{-1}}$ objects, followed by an inversion of such trend for more luminous, P$_{1.4\rm GHz}\simgt 10^{24.5}$ ${\rm W Hz^{-1} sr^{-1}}$, sources. 
On the other hand, Castignani et al. (2014) find no difference between the environmental properties of low-power and high-power radio AGN in their sample of $1\simlt z\simlt 2$ radio-selected AGN in the COSMOS region, while Malavasi et al. (2015), still working on radio-selected COSMOS sources in the redshift range $0\simlt z\simlt 2$ , find that radio fainter AGN prefer to inhabit overdense structures, at variance with the high radio-luminosity sources which show the same environmental properties of normal galaxies endowed with the same mass and redshift distributions. No connection between environment richness and nuclear radio luminosity is also claimed for CARLA AGN (Wyzalek et al. 2013).

As one of the most important results of this work, we find a {\it positive} trend between environment richness and radio luminosity in our sample of radio-selected AGN. Figure 4 shows the behaviour of radio AGN with luminosities respectively above and below the value of P$_{1.4\rm GHz}=10^{23.5}$ ${\rm W H^{-1} sr^{-1}}$. Despite the relatively poor statistics, it is nevertheless clear that there is a marked tendency for more radio-luminous AGN to appear within clusters ($38^{+15}_{-11}$\%) with respect to both fainter sources ($15\pm 3$\%) and the whole radio-selected AGN population ($\sim$18\%). This trend is reversed in underdense regions as only $24^{+13}_{-9}$\% of radio-bright AGN is associated with a field galaxy, while this is true for $42\pm 5$\% of the fainter ones. Once again, no difference is  found in the way radio AGN of different radio luminosities occupy filamentary structures.
Indeed, the above finding seems to highlight a monotonic trend which intrinsically depends on the radio luminosity of the sources as  the departure from the behaviour observed for the general population of radio-selected AGN is more and more visible as one considers AGN of higher and higher radio power (e.g. $26\pm 5$\% of  the sources brighter than P$_{1.4\rm GHz}=10^{23}$ ${\rm W Hz^{-1} sr^{-1}}$ are found within clusters, compared to the $\sim 18$\% of the whole radio AGN population and to the $\sim 38$\% of radio-selected AGN brighter than P$_{1.4\rm GHz}=10^{23.5}$ ${\rm W Hz^{-1} sr^{-1}}$).  

Another interesting finding on the way radio AGN occupy the cosmological volume is related to their emission properties in other wavebands.  
As already discussed in \S 2, radio-selected AGN in our sample have been further divided into those which also show signatures of AGN activity in the X-ray bands (X-ray emitters) and those whose Mid-Infrared SED indicates the presence of a buried AGN (MIR emitters).
By then associating these different sub-classes to the various cosmological environments identified within the COSMOS area (cfr Figure \ref{fig5}), we obtain that radio AGN which are also MIR emitters are preferentially found  in underdense regions. Indeed, 13 out of 18 (corresponding to $72^{+26}_{-20}$\% of the MIR-emitting sub-sample) appear in the field, while only 1 out of 18 (corresponding to $5^{+13}_{-3}$\%) is found within an overdensity. 
On the other hand,  the very few X-ray emitters all prefer rich environments and only result associated with clusters. Interestingly enough, both sub-classes seem to avoid filamentary structures. In Magliocchetti et al. (2018) we had already noticed that radio-selected AGN which also emit in the MIR seem to be younger and less massive than those radio-selected AGN which also emit in the X-ray. The present analysis then further suggests that these two sub-populations  behave very differently also when it comes to  their environmental properties, as the latter class seems to prefer overdense regions, while the former  is largely found amongst isolated galaxies. We also stress that our findings further confirm those obtained by Tasse et al. (2008), who indeed measure a MIR excess consistent with a hidden radiatively efficient active nucleus for their population of low-mass radio AGN, mostly observed to avoid overdensities.  In fact, in agreement with their results, we also find that MIR emitters are less massive and prefer large-scale underdense regions with respect to both the sub-class of X-ray emitters and the parent radio AGN population. \\
\begin{table}
\begin{center}
\caption{Properties of the AGN sample. Values between brackets indicate the relative percentages as illustrated in Figures 1-5. The quoted errors correspond to 1$\sigma$ Poisson uncertainties estimated following Gehrels (1986).}
\begin{tabular}{llllll}
 AGN & Total& Clusters& Filaments & Field\\
\hline
All& 218& 40 ($18^{+3}_{-3}$)& 91 ($42^{+4}_{-4}$)& 87 ($40^{+4}_{-4}$)\\
FIR-detected & 92& 11 ($12^{+5}_{-4}$)& 37 ($40^{+8}_{-7}$)&44 ($48^{+8}_{-7}$)\\
$P_{1.4 \rm GHz}>10^{23.5}$& 29& 11 ($38^{+15}_{-11}$)& 11 ($38^{+15}_{-11}$)&7 ($24^{+13}_{-9}$)\\
$P_{1.4 \rm GHz}<10^{23.5}$ & 189& 29 ($15^{+3}_{-3}$)& 80 ($42^{+5}_{-5}$)&80 ($42^{+5}_{-5}$)\\
$M_*\ge10^{11} M_\odot$ & 126& 30 ($24^{+5}_{-4}$)& 54 ($43^{+7}_{-6}$)&42 ($33^{+6}_{-5}$)\\
$M_*<10^{11} M_\odot$ & 92& 10 ($11^{+4}_{-3}$)& 37 ($40^{+8}_{-7}$)&45 ($49^{+8}_{-7}$)\\
$M_*\le 10^{10.4} M_\odot$ & 15& - & 5 ($33^{+22}_{-14}$)&10 ($67^{+28}_{-21}$)\\
MIR emitters & 18& 1 ($5^{+13}_{-3}$)& 4 ($22^{+18}_{-11}$)&13 ($72^{+26}_{-20}$)\\
X-ray emitters& 3& 3 ($100^{+0}_{-55}$)& -& -\\
\end{tabular}
\end{center}
\end{table}

\subsection{AGN distribution within clusters}
Darvish et al. (2017) also provide a classification for galaxies within overdense structures as either central or satellites: a {\it central} galaxy is identified as the most massive galaxy of each group/cluster, while all the other galaxies are classified as {\it satellites}. 
We use this further distinction to investigate our sample 
as a function of radio luminosity and stellar mass. This is done in Figure 6, which shows the distributions of radio luminosities and stellar masses for those radio-selected AGN residing in clusters. The first two plots of Figure 6 suggest that  there is very little (if any)  dependence of the positioning of a radio AGN within its cluster environment on its radio luminosity. This holds even for the most luminous, P$_{1.4\rm GHz}\ge10^{23.5}$ ${\rm W Hz^{-1} sr^{-1}}$, objects.  If we compare this result with that obtained earlier in this Section (cfr Figure 4), we can then conclude that (radio) AGN output is expected to be related to the preferred  location of the host galaxy within overdense structures, but is irrelevant to the favourite position (center or satellite) of the radio AGN  within its cluster environment. 

The situation changes if we investigate radio AGN of different stellar masses. Indeed, in this case not only we find that stellar mass is strongly related to the AGN choice to reside within cosmological regions of different richness (cfr Figures 2 and 3) but, although in this case the result might just reflect the well known phenomenon of mass segregation, it is also observed to strongly influence the preferred location of the AGN within its overdensity. The left-hand and right-hand panels of  Figure 6 clearly show  that this latter dependence is rather remarkable, as  almost all radio-selected AGN (10 out of 11) with $M_*>10^{11.3}$ M$_\odot$ coincide with  a cluster central galaxy, while 9 out of  10 of  those with $M_*<10^{11}$ M$_\odot$ are satellites.

\begin{table*}
\begin{center}
\caption{Properties of the considered sample. Column (1) presents the source identification number as taken from Darvish et al. (2017). Columns (2), (3) and (4) are the RA(2000), Dec(2000) and 1.4 GHz flux densities (measured in mJy units) of the radio sources from the Bondi et al. (2008) catalogue. Column (5) provides the decimal logarithm of the 1.4 GHz radio luminosity measured in ${\rm W Hz^{-1}sr^{-1}}$. Column (6) provides a classification of the sources as either AGN or star-forming galaxies from the work of Magliocchetti et al. (2018). Columns (7) and (8) present the photometric redshifts and stellar masses (in ${\rm M_\odot}$ units) of the sources from Laigle et al. (2016). Columns (9) and (10)  provide the environmental properties of the radio sources as taken from Darvish et al. (2017). Columns (11) and (12) provide $100 \mu$m and $160\mu$m fluxes as obtained from the PEP survey (Lutz et al. 2011) and measured in $\mu$Jy units, while columns (13) and (14)  are once again taken from the work of Magliocchetti et al. (2018)  and indicate whether the radio source has also been identified as an AGN respectively in the MIR and X-ray bands. The full table is available online.}
\begin{tabular}{llllllllllllll}
\hline
\hline
ID & RA& Dec&${\rm F_{1.4 GHz}}$&${\rm Log[P_{1.4 GHz}]}$&Type &  $z_{\rm phot}$ & ${\rm M_*}$&Environment&Position &$\rm {F_{100\mu m}}$  &
$\rm {F_{160\mu m}}$& MIR& Xray \\
\hline
(1)& (20&(30&(4)&(5)&(6)&(7)&(8)&(9)&(10)&(11)&(12)&(13)&(14)\\
\hline
  482487& 149.4192&   2.0292&  0.25& 22.70&AGN& 0.95& 11.30&   field  &  central & -& -&no&no\\
  372940& 149.4213&   1.8563& 45.62& 25.17&AGN& 0.98& 11.11& cluster  &satellite & -& -&no&yes \\
  509839& 149.4263&   2.0739& 20.49& 24.92&AGN& 1.08& 10.54&filament  & isolated &  20.94&  16.39&yes &no\\
  242528& 149.4301&   1.6475&  0.15& 22.06&SF & 0.52& 10.93& cluster  &satellite & -& -&no   &no   \\
  426476& 149.4371&   1.9413&  0.16& 21.87&SF & 0.42& 10.55&filament  &  central &  21.63&  25.97&no&no\\
  379494& 149.4470&   1.8647&  0.17& 22.06&SF & 0.50& 10.85&filament  &satellite &  25.53&  39.07&no&no\\
  467956& 149.4481&   2.0104&  0.62& 21.09&SF & 0.10& 10.13&filament  &  central & 184.96& 162.75&no   &no   \\
  420274& 149.4494&   1.9328&  0.17& 22.24&SF & 0.59& 10.89&filament  &  central &  20.55&  52.33&no   &no   \\
  416138& 149.4506&   1.9235&  0.16& 22.62&AGN& 0.94& 11.17&filament  &satellite & -& -&no   &no   \\
  752818& 149.4582&   2.4380&  0.22& 22.14&SF & 0.49&  9.86& cluster  &satellite &  29.88&  31.39&yes &no\\
  878956& 149.4597&   2.6299&  0.42& 23.02&AGN& 0.86& 10.24&filament  &satellite &   7.71& -&no   &no   \\

\end{tabular}
\end{center}
\end{table*}

\section  {Conclusions}
We have presented the environmental properties of $z\le 1.2$ radio-selected AGN obtained by pinpointing their location within the cosmic web of galaxies as identified in the COSMOS field by Darvish et al. (2017). The main results of our work can be summarised as follows:\\
1)  Radio-selected AGN  prefer rich environments. $18\pm 3$\% of them are in fact found within clusters, while the fraction of "normal" COSMOS galaxies within the same kind of structures is a factor 2 smaller (10.9\%; Darvish et al. 2017). The situation is reversed in the field ($40\pm 4$\% vs 48\%), while a similar percentage of radio-emitting AGN and normal galaxies is found to reside within filaments. The result  for a preponderance of radio-active AGN in overdense regions mirrors that obtained by Malavasi et al. (2015) also on the COSMOS field, despite the different criteria adopted by these latter authors to identify both {\it bona fide} AGN and overdensities.\\
2) Star-formation activity within the host galaxies of radio-active AGN does not seem to be related to the large-scale environment of the galaxy itself. 
In fact, radio-selected AGN associated with ongoing star formation are found within environments which are indistinguishable from those inhabited by  the parent radio AGN population.\\
3) Both radio luminosity and stellar content of the host galaxy seem to affect the location of the radio AGN  within different regions of the cosmic web. In particular, we find that $24^{+5}_{-4}$\% of those with $M_*\ge10^{11}$ M$_\odot$ are located in clusters, while this is only true  for $11^{+4}_{-3}$\% of the radio AGN with $M_*<10^{11}$ M$_\odot$. Such a deficit of low-mass radio-emitting AGN within overdense/cluster-like structures is balanced off in the field where we find more low-mass sources ($49^{+8}_{-7}$\%) than more massive ones ($33^{+6}_{-5}$\%). No difference in the environmental properties of the two sub-populations is instead observed within filaments.\\
We stress that the above results may not be solely due to the well known phenomenon of mass segregation certainly present also in our case, as we find $(1.5-2)\sigma$ differences in the way radio-selected AGN and normal galaxies with the same stellar mass content occupy the various cosmological structures. These differences can almost certainly be ascribed to the presence of nuclear radio activity, and our results suggest that such an activity tends to be favoured either in large,  $M_*\simgt10^{11}$ M$_\odot$, stellar mass systems residing within overdense structures or in smaller ($M_*/M_\odot\simlt 10^{10.4}$), more isolated galaxies. \\
4) Even more striking is the  dependence of the environment of radio-selected AGN on their radio luminosity: indeed, $38^{+15}_{-11}$\% of the sources more powerful than P$_{1.4\rm GHz}=10^{23.5}$ $\rm WHz^{-1} sr^{-1}$ are found within clusters, while this is true for only $15\pm 3$\% of those with P$_{1.4\rm GHz}<10^{23.5}$ $\rm WHz^{-1} sr^{-1}$.
At the same time, we find that 
only $24^{+13}_{-9}$\% of the radio-luminous, P$_{1.4\rm GHz}>10^{23.5}$ $\rm WHz^{-1} sr^{-1}$, AGN  are associated with isolated galaxies, while the percentage for  the fainter ones is sensibly higher ($42\pm 5$\%).\\
5) AGN emission at wavelengths other than radio also seems to be connected with the favourite environment of radio-active AGN. In fact, while radio AGN which are also active in the MIR are mainly found as isolated galaxies ($72^{+26}_{-20}$\% in the field vs $5^{+13}_{-3}$\%  in clusters), the opposite seems to be true for the sub-population of radio-selected X-ray emitters, as the few which are present in our sample all reside within cluster-like structures.\\
The existence of a class of radio-selected AGN which also emit in the MIR and are hosted by  low-mass (M$_*\simlt 10^{10.5}$ M$_\odot$) galaxies which preferentially reside in underdense  regions might suggest different mechanisms triggering nuclear radio activity, e.g. galaxy-galaxy interaction in low-mass galaxies and gas accretion from the surrounding cluster in high-mass galaxies (cfr Tasse et al. 2008). Further work is needed to test and eventually confirm this intriguing scenario.\\
6) Lastly, we studied the distribution of radio-active AGN within cluster-like structures as a function of both radio luminosity and stellar content of the hosts. Our results indicate that while there is a strong dependence on the stellar mass as virtually all AGN hosts associated with cluster satellite galaxies have $M_*<10^{11.3}$ M$_\odot$ and all those which coincide with a cluster central galaxy have $M_*>10^{11}$ M$_\odot$, the specific location of a radio AGN within the cluster environment is only marginally dependent on its radio luminosity. Interestingly, the above results mirror those presented by Allevato et al. (2012) for a sample of $z<1$ X-ray selected AGN.

What emerges from our work is therefore a scenario whereby the physical properties of the sources on sub-pc/pc (radio luminosity and/or AGN emission at different wavelengths) and kpc (stellar mass) scales strongly influence the large-scale structure behaviour  of the AGN itself, up to the point of possibly determining its favourite environment. On the other hand, the precise location of radio-emitting AGN within overdense structures only seems to be related to the stellar mass associated with the AGN host. \\ 

\noindent {\bf Acknowledgements}
MM  wishes to thank G. Caprini, I. Codispoti, F.R. Fuxa, D. Laurenti, C. Levi, C. Ratti and R. Tomassetti for help and support during the completion of this work. R. Masetti, M. Salgarello and L. Barone for giving me the chance to complete it. 
We also wish to thank the referee Dr Heinz Andernach for valuable comments which have largely improved the paper.


\begin{thebibliography}{}
\bibitem[\protect\citeauthoryear{Allevato}2012]{Allevato}
Allevato V. et al., 2012, ApJ, 758, 47
\bibitem[\protect\citeauthoryear{Best}2004]{Best}
Best P. N., 2004, MNRAS, 351, 70
\bibitem[\protect\citeauthoryear{Best1}2007]{Best1}
Best P.N., von der Linden A., Kauffmann G., Heckman T.M.,  Kaiser C.R., 2007, MNRAS, 379, 894
\bibitem[\protect\citeauthoryear{Bondi}2008]{Bondi}
Bondi M., Ciliegi P., Schinnerer E., Smolcic V., Jahke K., Carilli C., Zamorani G., 2008, ApJ, 681, 1129
\bibitem[\protect\citeauthoryear{Brusa}2010]{Brusa}
Brusa M. et al., 2010, ApJ, 716, 348
\bibitem[\protect\citeauthoryear{Castignani}2014]{Castignani}
Castignani G., Chiaberge M., Celotti A., Norman C., De Zotti G., 2014, ApJ, 792, 114 
\bibitem[\protect\citename{Chang}2017]{Chang}
Chang Y. et al., 2017, ApJS, 233, 19
\bibitem[\protect\citename{Darvish}2017]{Dar}
Darvish B. et al., 2017, ApJ, 837, 16
\bibitem[\protect\citename{Donoso}2010]{Donoso}
Donoso E., Cheng L., Kauffmann G., Best P.N., Heckman T.M., 2010, MNRAS, 407, 1078
\bibitem[\protect\citename{Donley}2012]{Don}
Donley J. L. et al., 2012, ApJ, 748, 142
\bibitem[\protect\citename{fine}2011]{fine} 
Fine S., Shanks T., Nikoloudakis N., Sawangwit U., 2011, MNRAS, 418, 2251
\bibitem[\protect\citename{geh}1986]{geh} 
Gehrels N., 1986, ApJ, 303, 336
\bibitem[\protect\citename{Hatch}2014]{Hatch}
Hatch N.A. et al. 2014, MNRAS, 445, 280
\bibitem[\protect\citename{Ilbert}2013]{Ilbert}
Ilbert O., et al., 2013, A\&A, 556, A55
\bibitem[\protect\citename{Laigle}2016]{Lai}
Laigle C. et al. 2016, ApJS, 224, 24
\bibitem[\protect\citename{Lind}2014a]{Lind}
Lindsay S.N. et al., 2014a, MNRAS, 440, 1527
\bibitem[\protect\citename{Lind1}2014b]{Lin1}
Lindsay S.N., Jarvis M.J.,  McAlpine K., 2014b, MNRAS, 440, 2332
\bibitem[\protect\citename{Lutz}2011]{Lutz}
Lutz D. et al., 2011, A\&A, 532, A90
\bibitem[\protect\citename{Maglio1}2002]{Maglio1}
Magliocchetti M. et al.  2002, MNRAS, 333, 100
\bibitem[\protect\citename{Maglio2}2004]{Maglio2}
Magliocchetti M. et al.  2004, MNRAS, 350, 1485
\bibitem[\protect\citename{Maglio12}2007]{Maglio12}
Magliocchetti M., Br\"uggen M., 2007, MNRAS, 379, 260
\bibitem[\protect\citename{Maglio3}2014]{Maglio3}
Magliocchetti M. et al., 2014, MNRAS, 442, 682
\bibitem[\protect\citename{Maglio4}2016]{Maglio4}
Magliocchetti M., Lutz D., Santini P., Salvato M., Popesso P., Berta S., Pozzi F., 2016, MNRAS, 456, 431
\bibitem[\protect\citename{Maglio6}2017]{Maglio6}
Magliocchetti M., Popesso P., Brusa M., Salvato M., Laigle C., McCracken H.J., Ilbert I., 2017, MNRAS, 464, 3271
\bibitem[\protect\citename{Maglio5}2018]{Maglio5}
Magliocchetti M., Popesso P., Brusa M., Salvato M., 2018, MNRAS, 473, 2493
\bibitem[\protect\citename{Malavasi}2015]{Malavasi}
Malavasi N., Bardelli S., Ciliegi P., Ilbert O., Pozzetti L., Zucca E., 2015, A\&A, 576, A101
\bibitem[\protect\citename{Marchesi}2016]{Marchesi}
Marchesi S. et al., 2016, 830, 100
\bibitem[\protect\citename{McAlpine}2012]{McAlpine}
McAlpine K., Jarvis M.J., Bonfield D.G., 2013, MNRAS, 436, 1084
\bibitem[\protect\citename{Padovani} 2015]{Padovani}
Padovani P., Bonzini M., Kellermann K., Mainieri V., Miller N., Tozzi P., 2015, MNRAS, 452, 1263
\bibitem[\protect\citename{Porci} 2004]{Porci}
Porciani C., Magliocchetti M., Norberg P., 2004, MNRAS, 355, 1010
\bibitem[\protect\citename{Retana} 2017]{Retana}
Retana-Montenegro E., R\"ottgering H.J.A., 2017, A\&A, 600, A97
\bibitem[\protect\citeauthoryear{Schin}2007]{Schin2}
Schinnerer E. et al.., 2007, ApJS, 172, 46
\bibitem[\protect\citeauthoryear{Scov}2007]{Scov}
Scoville N. et al., 2007, ApJS, 172, 1
\bibitem[\protect\citename{Shen1} 2017]{shen1}
Shen L. et al., 2017, MNRAS, 472, 998
\bibitem[\protect\citename{Shen} 2009]{shen}
Shen Y. et al., 2009, ApJ, 697, 1656
\bibitem[\protect\citeauthoryear{Tasse}2008]{Tasse}
Tasse C., Best P.N., R\"ottgering H., Le Borgne D., 2008, A\&A, 490, 893
\bibitem[\protect\citename{Wake}2008]{Wake}
Wake D.A., Croom S.M., Sadler E.M., Johnston H.M., 2008, MNRAS, 391, 1674
\bibitem[\protect\citename{Wylezalek}2014]{Wylezalek}
Wylezalek D. et al., 2013, ApJ, 769, 79
\end{thebibliography}
\end{document}